\documentclass[conference,a4paper]{IEEEtran}
\usepackage{xcolor}
\usepackage{balance}
\usepackage{amsmath,amssymb,amsfonts}
\usepackage{subcaption}
\usepackage{algorithm}
\usepackage{algpseudocode}

\ifCLASSINFOpdf
  \usepackage[pdftex]{graphicx}
\else
  \usepackage[dvips]{graphicx}
\fi
\ifCLASSOPTIONcompsoc
 \usepackage[caption=false,font=normalsize,labelfont=sf,textfont=sf]{subfig}
\else
 \usepackage[caption=false,font=footnotesize]{subfig}
\fi
\begin{document}
%
\title{Supervised Learning Based Real-Time Adaptive Beamforming On-board Multibeam Satellites}

\author{\IEEEauthorblockN{
Flor Ortiz,   
Juan A. Vasquez-Peralvo,   
Jorge Querol,    
Eva Lagunas,      
Jorge L. Gonz\'alez Rios,\\
Marcele O. K. Mendon\c{c}a,
Luis Garces,
Victor Monzon Baeza,
Symeon Chatzinotas
}                                     
\IEEEauthorblockA{\IEEEauthorrefmark{1}
\textit{Interdisciplinary Centre for Security, Reliability and Trust (SnT)},\\ \textit{University of Luxembourg}, Luxembourg, Luxembourg, flor.ortiz@uni.lu}
}


\maketitle

\begin{abstract}
Satellite communications (SatCom) are crucial for global connectivity, especially in the era of emerging technologies like 6G and narrowing the digital divide. Traditional SatCom systems struggle with efficient resource management due to static multibeam configurations, hindering quality of service (QoS) amidst dynamic traffic demands. This paper introduces an innovative solution - real-time adaptive beamforming on multibeam satellites with software-defined payloads in geostationary orbit (GEO). Utilizing a Direct Radiating Array (DRA) with circular polarization in the 17.7 - 20.2 GHz band, the paper outlines DRA design and a supervised learning-based algorithm for on-board beamforming. This adaptive approach not only meets precise beam projection needs but also dynamically adjusts beamwidth, minimizes sidelobe levels (SLL), and optimizes effective isotropic radiated power (EIRP).
\end{abstract}

\vskip0.5\baselineskip

%

\section{Introduction}
Satellite communications (SatCom) is a fundamental pillar of modern global connectivity, providing the means to bridge the digital divide and offering ubiquitous coverage in an increasingly connected world. The integration of terrestrial systems such as 6G further underlines the importance of SatCom as it continues to facilitate communication on a global scale \cite{giordani2020satellite}. However, the increased data traffic on SatCom systems poses a considerable challenge: effectively managing radio resource allocation while meeting stringent quality of service (QoS) requirements remains a formidable task \cite{cornejo2022method}.

Traditionally, SatCom systems have relied on static multibeam configurations with fixed bandwidth and power allocations. However, these configurations fall short of adapting to the dynamic nature of today's traffic demands, leading to inefficient resource utilization and potential service degradation. Recognizing the temporal and spatial variations in demand, software-defined payloads have emerged as a revolutionary solution. These payloads offer unprecedented flexibility and adaptability in radio resource management (RRM) for SatCom \cite{angeletti2021heuristic}.

While software-defined payloads are very promising, their effective utilization requires advanced RRM techniques to optimize resource allocation in real-time. A crucial facet of RRM in SatCom is the adaptive beamwidth, power, and pointing through beamforming control. Conventional optimization-based approaches, while theoretically sound, often lack the computational efficiency and adaptability needed to cope with the diverse and dynamic traffic patterns encountered in SatCom systems \cite{lim2019payload} . 

Recent studies have proposed schemes to enhance spectral efficiency and user fairness in multi-beam satellite systems through robust beamforming and non-orthogonal multiple access, considering imperfect channel information among terminals \cite{9610022}. There's also a shift towards adaptive multibeam planning and demand-based footprinting to cater to dynamic traffic demands, especially in remote areas \cite{9469916}. While technological advances have spurred the use of all-digital phased arrays, the high computational cost of adaptive beamforming remains a challenge, with some research exploring neural networks for real-time scenarios, albeit without fully addressing service area traffic demands and SatCom system constraints \cite{9020740}. 

This paper explores two novel approaches to on-board real-time adaptive beamforming based on supervised learning. Specifically, the use of a Direct Radiating Array (DRA) operating in circular polarization within the frequency band 17.7 - 20.2 GHz is proposed. In addition to addressing the beam requirements, this research delves into other vital parameters such as beamwidth in the azimuthal and elevation planes, sidelobe level (SLL) control, and effective isotropic radiated power (EIRP) control. 

\section{System Model and Problem Statement}
In this section, we present the system model and outline the optimization problem for our GEO satellite system. Our system comprises a single multibeam satellite that serves a wide Earth region with $B$ spot-beams. The focus lies on the forward user link, with $K$ single-antenna user terminals (UTs) distributed across the satellite's coverage area. The satellite's payload features versatile power management using a traveling-wave tube amplifier (TWTA) with adaptive input back-off (IBO) and beamwidth control using a DRA. Frequency reuse strategies are implemented to mitigate co-channel beam interference.

The RRM objective is to efficiently allocate available resources, minimizing the discrepancy between the offered capacity $C_{\tau}^b$ and the requested capacity $R_{\tau}^b$ in the $b$-th beam during time slot $\tau$. The offered capacity $C_{\tau}^b$ [bps] in the $b$-th beam at time slot $\tau$ is computed as
\begin{equation}
    C_{\tau}^b = \mathrm{W}_{\tau}^{b} \cdot \mathrm{\kappa}_{\tau}^b,
    \label{offcap_eq}
\end{equation}
where $\mathrm{\kappa}_{\tau}^b$ [bps/Hz] represents the spectral efficiency of the selected modulation and coding scheme, and $\mathrm{W}_{\tau}^{b}$ [Hz] denotes the allocated bandwidth for the $b$-th beam. The spectral efficiency $\mathrm{\kappa}_{\tau}^b$ relies on the Carrier-to-Interference-plus-Noise Ratio (CINR) $\gamma^b_{\tau}$ of the $b$-th beam at slot $\tau$, governed by a mapping function $\mathrm{\kappa}_b = f(\gamma_b)$.

The CINR $\gamma^b_{\tau}$ depends on the power and beamwidth allocated to the $b$-th beam, determined by the DRA. The channel gain for beam $b$ is computed according to a standard model:
\begin{equation}
    |h^b|^2 = \frac{G_{\text{SAT}}(\theta_{b}) G_{\text{RX,max}}}{\left(4 \pi d^b/\lambda\right)^{2}L^b},
    \label{eq}
\end{equation}
with $d^b$ representing the distance between the satellite and the center of the $b$-th beam on the ground, $\lambda$ indicating the wavelength, $L^b$ accounting for shadowing and atmospheric gas losses, and $\theta_{b}$ denoting the satellite off-boresight transmit angle towards the beams. $G_{\text{SAT}}(\theta)$ and $G_{\text{RX,max}}$ denote the satellite antenna gain at a specific off-boresight angle and the maximum receive antenna gain, respectively.



Our system incorporates a DRA with $N \times N$ radiating elements to control the power, beamwidth, Equivalent Isotropic Radiated Power (EIRP), which depends on the power and beamwidth, Side Lobe Level (SLL), and pointing per beam depending on traffic requirements. To determine the number of antenna elements, we consider gain requirements, beam solid angle, satellite position, altitude, and coverage area. 


To optimize these parameters, a beamforming cost function is defined to optimize the weight matrix, $W_{p\times q}^B$. This function minimizes the error in required beamwidth, SLL, and EIRP. This optimization problem enables the adjustment of antenna parameters to achieve the desired performance while accounting for practical constraints \cite{vasquez2023flexible}:
\begin{equation}
\label{eq:CostFunction}
\min_{W_{p\times q}^B} \left( Z_1(W_{p\times q}^B) + Z_2(W_{p\times q}^B) + Z_3(W_{p\times q}^B) \right),
\end{equation}
where:
\begin{equation*}
\label{eq:Eq2bF}
    \left\{
    \begin{aligned}
        & Z_1 =  \Bigg( \frac{|\mathrm{\theta_{-3dB_{Az_c}}^b}(W_{p\times q}^B)-\mathrm{\theta_{-3dB_{Az_o}}^b}|}{\mathrm{\theta_{-3dB_{Az_o}}^b}}+ \\ 
        & \hspace{1cm} \frac{|\mathrm{\theta_{-3dB_{El_c}}^b}(W_{p\times q}^B)-\mathrm{\theta_{-3dB_{El_o}}^b}|}{\mathrm{\theta_{-3dB_{El_o}}^b}} \Bigg)k_1\\  
        & Z_2 = \Bigg(\frac{|\mathrm{SLL_{Az_c}^b}(W_{p\times q}^B)-\mathrm{SLL_{Az_o}^b}|}{\mathrm{SLL{Az_o}^b}}+ \\
        & \hspace{1cm} \frac{|\mathrm{SLL_{El_c}^b}(W_{p\times q}^B)-\mathrm{SLL_{El_o}^b}|}{\mathrm{SLL{El_o}^b}}\Bigg)k_2\\
        &  Z_3 = \left(\frac{\mathrm{EIRP_c^b}(W_{p\times q}^B)-\mathrm{EIRP_o^b}}{\mathrm{EIRP_o^b}}\right)k_3. \\
    \end{aligned}
    \right.
\end{equation*}
where $Z_1$, $Z_2$, and $Z_3$ represent sub-objectives that quantify the discrepancies in the required beamwidth, SLL, and EIRP, defined by sub-index \textit{o} respectively and the real defined by sub-index \textit{c}. The weights $k_1$, $k_2$, and $k_3$ allow for fine-tuning the importance of each sub-objective in the optimization process.

\section{Antenna Design and Training Data Generation}
\subsection{Antenna Design}
For GEO missions, efficient antenna design is crucial due to substantial free space losses and stringent constraints on permissible losses. To address these challenges, we have developed an open-ended waveguide antenna as the unit cell antenna for this scenario, as explained in \cite{vasquez2023flexible}. This antenna design offers significantly lower losses than alternative solutions like patch antennas or dielectric-based antennas. The antenna consists of three main components: the open-ended waveguide itself, the groove polarizer responsible for circular polarization, and the rectangular-to-circular transition for connecting to the distribution network.

%


Simulation results, confirm that the antenna meets the requirements. It radiates within the required frequency band, emitting Left-Hand Circular Polarized (LHCP) waves with minimal cross-polarization. This is evident from the $S_{11}$ parameter, below -10 dB, and the axial ratio measuring less than 3 dB within the intended frequency range \cite{vasquez2023flexible}.

We consider the coverage area over the Earth's surface to determine the number of elements required for the array antenna. In the context of Very High Throughput satellite missions, where small areas need to be covered to avoid channel link saturation, we set a minimum beam diameter of 260 km or a coverage area $A_c = 53093 \, \text{km}^2$ when the satellite is at nadir. It's worth noting that the coverage area can be adjusted based on specific beam requirements.

We then calculate the 3 dB antenna beamwidth $\theta_{-3dB}$ required to illuminate this area. Using geometric relations and the satellite's location, we estimate $\theta_{-3dB}$, which in this case is found to be $0.41^\circ$.

Next, we determine the number of antenna elements per dimension, denoted as $N_x$ and $N_y$, using the formula:
\begin{equation}
    \label{Eq.NxNy}
    N_x = N_y = \frac{0.886 \lambda_0}{\eta \theta_{-3dB}d}
\end{equation}
where $\lambda_0$ is the free space wavelength, $\eta$ is the antenna efficiency, and $d$ is the antenna inter-element spacing. For an operational frequency of $f_0 =$ 19 GHz and assuming maximum efficiency, the total number of elements is determined to be $N_x = N_y = 144$.

However, this number of elements is impractical regarding space, cost, and power requirements, as each element requires an RF chain. To address this, subarrays are introduced. The number of subarray elements is determined to avoid grating lobes intersecting the antenna's Field of View (FoV) over the Earth's surface resulting in an element spacing of $d = 3.5 \lambda_0$. With this spacing, subarrays of 4$\times$4 elements are employed, resulting in 36$\times$36 RF chains.

\subsection{Training Data Generation}
\begin{figure*}[!htp]
     \centering
        \begin{subfigure}[b]{0.35\textwidth}
         \centering
         \includegraphics[trim=80 30 25 80,clip,width=\textwidth]{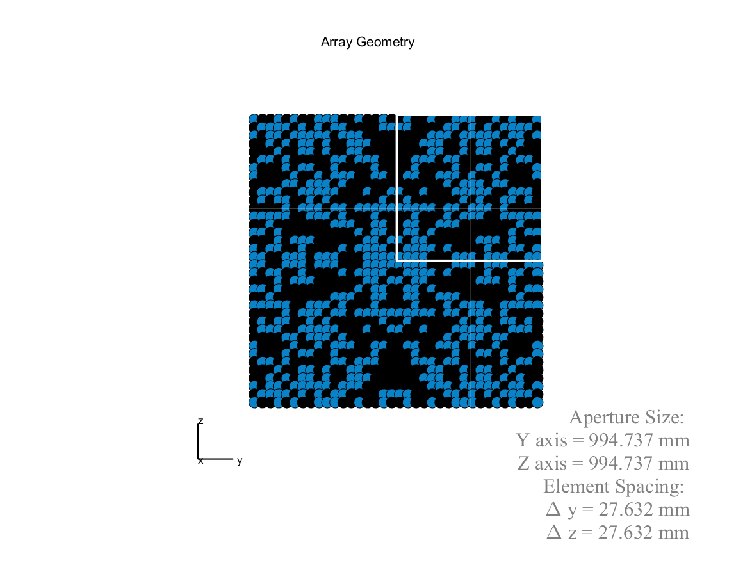}
         \caption{}
         \label{fig:Fig2a}
     \end{subfigure}
     \hfil
     \begin{subfigure}[b]{0.35\textwidth}
         \centering
         \includegraphics[trim=80 25 25 80,clip,width=\textwidth]{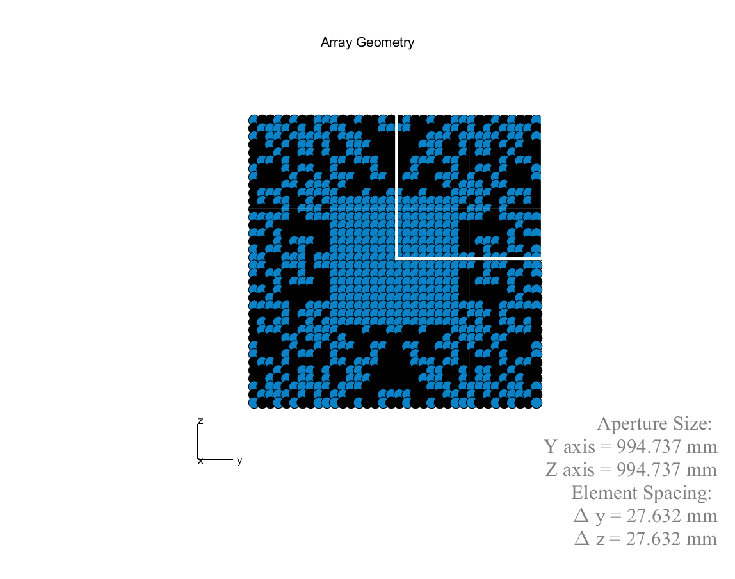}
         \caption{}
         \label{fig:Fig2b}
     \end{subfigure}
     \hfil
     \begin{subfigure}[b]{0.42\textwidth}
         \centering
         \includegraphics[width=\textwidth]{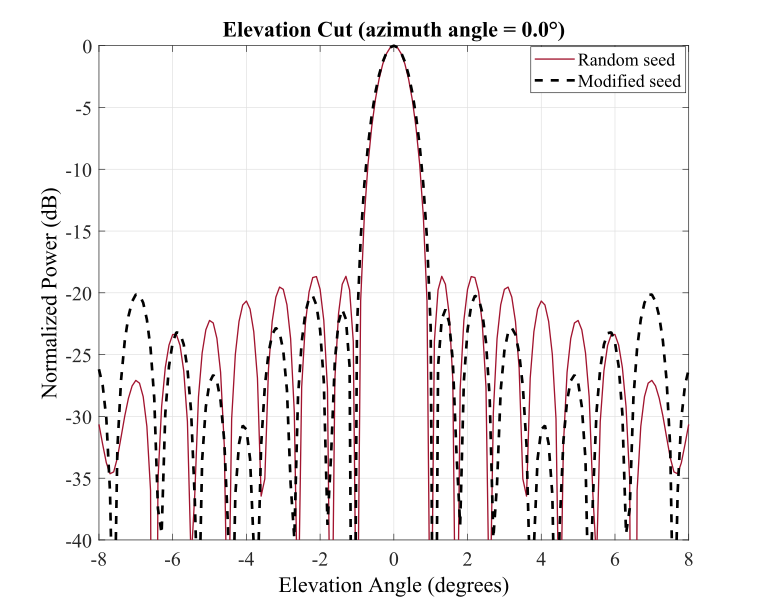}
         \caption{}
         \label{fig:Fig2c}
     \end{subfigure}
     
         \caption{a) Random elements. b) Mixed, fixed at the center, and random at the borders. c) Radiation pattern azimuth cut for the two weight matrices.}
        \label{fig:antenna}
\end{figure*}

The approach chosen in this study is array thinning, which consists of selectively activating and deactivating elements to form the desired antenna beam, as illustrated in Figure \ref{fig:antenna}. Array thinning allows precise control of the beamwidth, uniform power distribution among the elements in multi-beam scenarios and orientation of the radiation pattern in the desired direction by progressive phase shifting.

To create the training data for this antenna array, a Genetic Algorithm (GA) was employed. The goal of the GA was to find the optimal set of active antenna elements while respecting specific constraints, such as beamwidth, SLL and EIRP.

The GA optimization process iteratively refines the antenna element configurations, starting from an initial configuration and continuing until convergence or until a predefined maximum iteration limit is reached. It considers several performance parameters simultaneously, such as beamwidth, SLL and EIRP, making it particularly suitable for optimizing complex antenna systems such as the distributed reflector antenna.

The resulting training dataset consists of 174,203 samples, each representing different antenna element configurations that meet the defined constraints. This dataset serves as a reference for evaluating antenna performance and training machine learning models in the later sections of this paper.

The main purpose of this database is to establish correlations between beamforming array weights and critical system parameters, including beamwidth, SLL, and EIRP. For more detailed information on the generated database, including access to the dataset itself, see \cite{fnr-smartspace}, where it is openly available.

\section{Supervised Learning for Adaptive Beamforming}
The design of our beamforming matrix allows us to divide the 36-element matrix into four distinct 18-element sections for more detail see \cite{vasquez2023flexible}. Consequently, our goal is reduced to predicting the $18 \times 18$ matrix, comprising a total of 324 elements.

For this study. we present two approaches based on supervised learning: approach 1, based on Multi-Label Classification Neural Network, and approach 2 based on Clustering and Classification Neural Network.

Approach 1 employs a multi-label classification neural network for beamforming, as shown in Algorithm 1. This method is designed to predict the activation state of individual elements within the beamforming matrix. The primary goal is determining which elements should be active (assigned a value of 1) and which should be inactive (assigned a value of 0) based on beam-related input features.

The neural network takes eight beam-related input features, such as beam width in azimuth, beam width in elevation, minimum SLL in azimuth, minimum SLL in elevation, equivalent isotropic radiated power ($EIRP_b$), azimuth, elevation, and number of elements. The output layer consists of 324 units, corresponding to the 324 elements of the $18times 18$ section of the beamforming matrix. Each unit in the output layer uses a sigmoid activation function, which produces values between 0 and 1, indicating the probability that each element is active. 
    
The loss function used is the binary cross-entropy, which calculates the error of each element individually and then aggregates them to evaluate the overall model performance. Model performance is evaluated by the multi-label accuracy or the $f1_{score}$, which measures the percentage of correct element activations predicted for each input.

The neural network architecture for Approach 1 is constructed accordingly, with input, hidden, and output layers. Once trained, this model can predict the activation state of the elements in the beamforming matrix based on the characteristics of the input beams.

Approach 2 employs a two-step process involving clustering and classification to determine the appropriate beamforming matrix as explained in Algorithm 2. Initially, a K-means-based clustering algorithm is used to group similar sets of input variables related to the beamforming matrix design. This clustering operation transforms the problem into a binary classification scenario, where each class corresponds to a specific predefined beamforming matrix.

The neural network used for classification is designed to learn and solve this binary classification task by analyzing the characteristics of the input variables and assigning them to the appropriate pre-trained matrix. This approach aims to optimize the selection of beamforming matrices based on the input variables, improving the system's overall performance.

\begin{algorithm}[!htp]
  \caption{Multi-Label Classification Neural Network (Approach 1)}
  \begin{algorithmic}[1]
    \Procedure{MultiLabelBeamforming}{$\text{Input Features}$}
    \State Initialize the neural network architecture
    \State Define input layer with 8 neurons, one for each input feature
    \State Define three hidden layers with ReLU activation functions
    \State Define the output layer with 324 units and sigmoid activation
    \State Compile the neural network using binary cross-entropy loss
    \State Train the model using labeled data
    \State Evaluate the model's performance using multi-label accuracy or $f1_{score}$
    \State Predict element activations for new input features
    \State \textbf{Output:} Predicted element activations
    \EndProcedure
  \end{algorithmic}
\end{algorithm}

\begin{algorithm}[!htp]
  \caption{Clustering and Classification Neural Network (Approach 2)}
  \begin{algorithmic}[1]
    \Procedure{ClusterAndClassifyBeamforming}{$\text{Input Features}$}
    \State Cluster input features using K-means algorithm
    \State Determine the optimal number of clusters based on analysis
    \State Assign each input data point to one of the clusters
    \State Define a Classification Neural Network
    \State Design input, hidden, and output layers for the neural network
    \State Train the model using labeled data (clusters as labels)
    \State Evaluate the model's performance using accuracy
    \State Predict the cluster (pre-defined matrix) for new input features
    \State \textbf{Output:} Predicted cluster (pre-defined matrix)
    \EndProcedure
  \end{algorithmic}
\end{algorithm}

\section{Numerical Results}
The neural networks are trained offline, so they are used only for onboard satellite inference which drastically decreases the execution time, as explained in \cite{https://doi.org/10.1002/sat.1422}. Both approaches are suitable for real-time beamforming adaptation due to the speed with which $W_{p\times q}^B$ can be computed as a function of requirements. To evaluate the performance of our two approaches, we separated our database into three sets, the training set, the test set and the validation set with a distribution of 70\%, 15\% and 15\% respectively. The results obtained with the validation set are compared with the GA for the same input conditions and we define six Key Performance Indicators (KPI) to evaluate their performance in terms of various metrics, it is important to note that we have taken the average obtained during all the tests:

\begin{enumerate}
    \item Matched EIRP (KPI 1):
    \[
    KPI_1 = \left(1 - \frac{|EIRP_y - EIRP_x|}{EIRP_x}\right) \times 100
    \]
    where $EIRP_x$ is the desired EIRP, and $EIRP_y$ is the obtained EIRP with the evaluated approach.

    \item Matched Beamwidth (KPI 2):
    \[
    KPI_2 = \left(1 - \frac{|\theta_y - \theta_x|}{\theta_x}\right) \times 100
    \]
    where $\theta_x$ is the desired beamwidth, and $\theta_y$ is the obtained beamwidth with the evaluated approach.

    \item Matched SLL (KPI 3):
    \[
    KPI_3 = \left(1 - \frac{|SLL_y - SLL_x|}{SLL_x}\right) \times 100
    \]
    where $SLL_x$ is the desired SLL, and $SLL_y$ is the obtained SLL with the evaluated approach.

    \item Matched Demand (KPI 4):
    \[
    KPI_4 = \left(1 - \frac{|C^b - D^b|}{D^b}\right) \times 100
    \]
    where $D^b$ is the demand, and $C^b$ is the obtained capacity with the evaluated approach.

    \item F1 Score of the Algorithm (KPI 5):
    \[
    KPI_5 = \left(1 - \frac{2 \cdot (Precision \cdot Recall)}{Precision + Recall}\right) \times 100
    \]
    where Precision measures the model's ability not to misclassify negative instances, and Recall measures the model's ability to find all positive instances.

    \item Algorithm Speed (KPI 6):
    \[
    KPI_6 = \left(1 - \frac{execution\_time}{reference\_time}\right) \times 100
    \]
\end{enumerate}

\begin{figure}[h]
    \centering
    \includegraphics[width=0.52\textwidth]{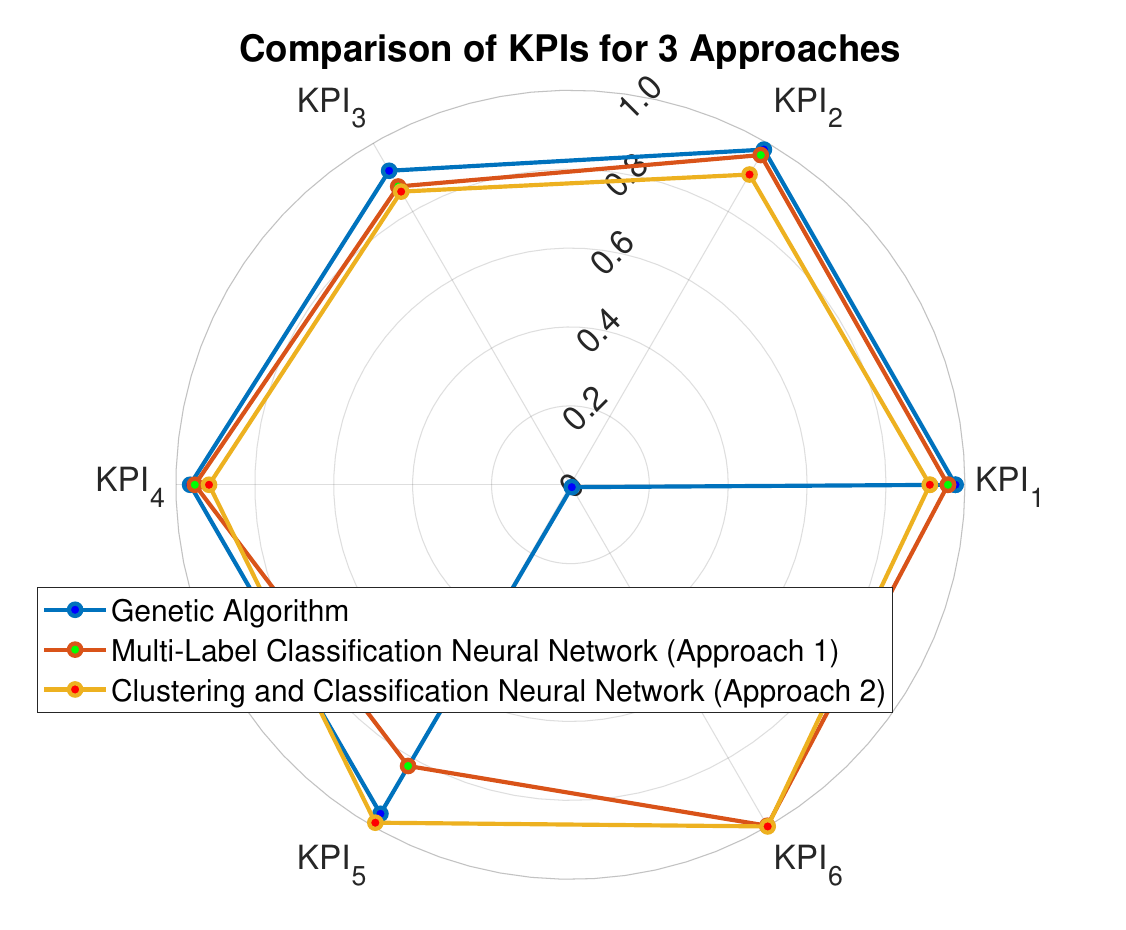}
    \caption{Average KPIs for the three approaches}
    \label{fig:kpis}
\end{figure}

Figure \ref{fig:kpis} shows the average KPIs obtained for the three approaches: Genetic Algorithm, Multi-Label Classification Neural Network (Approach 1), and Clustering and Classification Neural Network (Approach 2) after running and testing the three approaches iteratively with a sample size greater than 50,000 samples.

The Genetic Algorithm outperforms almost all the evaluated KPIs. However, the time required for execution and obtaining the beamforming matrix makes this algorithm unsuitable for real-time applications. On the other hand, both supervised learning-based approaches significantly reduce the execution time by more than a thousand times. This is because after training the ML models, they can be used for inference with almost immediate response times, making them suitable for real-time adaptation. Additionally, both ML-based approaches consistently maintain system performance above 90

Regarding classification metrics, Approach 2 outperforms Approach 1 because binary classification is typically less complex than multi-label classification. However, in terms of overall system performance, Approach 1 performs better than Approach 2.

\section{Conclusions}
We present two novel approaches to adaptive beamforming in satellite communications systems and compare them to the traditional GA. Our goal was to explore more efficient and real-time alternatives to GA, which, although effective, can be computationally expensive. Both approaches offer viable alternatives to GA for adaptive beamforming in satellite communication systems. These approaches not only reduce execution times but also maintain high system performance. 

Selection between the two approaches should be based on the specific requirements of the application, taking into account the trade-off between classification complexity and overall system performance. As a result, these findings open up new possibilities for implementing efficient, real-time adaptive beamforming in wireless communication systems.

Future lines of research may include further optimizations of these approaches, exploration of hybrid models, or incorporation of additional features to improve performance and versatility.


\section*{Acknowledgment}
This work was supported by the European Space Agency
(ESA) funded under Contract No. 4000134522/21/NL/FGL
named “Satellite Signal Processing Techniques using a Commercial Off-The-Shelf AI Chipset (SPAICE)”. Please note that the views of the authors of this paper do not necessarily reflect the views of the ESA. Furthermore, this work was partially supported by the Luxembourg National Research Fund (FNR) under the project SmartSpace (C21/IS/16193290).

\bibliographystyle{IEEEtran}
\bibliography{references.bib}

\end{document}